\documentclass[aps,amsmath,twocolumn,amssymb,floatfix,showpacs]{revtex4}
\usepackage[dvips]{graphics}
\usepackage{color}
\definecolor{gold}{rgb}{0.85,0.66,0}
\definecolor{dgreen}{rgb}{0.3,0.6,0.2}

\begin{document}

\title{\textcolor{blue}{Engineering flat electronic bands in quasiperiodic and 
fractal loop geometries}}
\author{Atanu Nandy}
\email[\textbf{E-mail:} atanunandy1989@gmail.com]{}
\author{Arunava Chakrabarti}
\email[\textbf{E-mail:} arunava\_chakrabarti@yahoo.co.in]{}
\affiliation{Department of Physics, University of Kalyani, Kalyani,
West Bengal - 741 235, India}
\begin{abstract}
Exact construction of one electron eigenstates with 
flat, non-dispersive bands, and localized over clusters of 
various sizes is reported for a class of quasi-one dimensional looped 
networks. Quasiperiodic Fibonacci and Berker 
fractal geometries are embedded in the 
arms of the loop threaded by a uniform magnetic flux. We 
work out an analytical scheme to unravel the localized single 
particle states pinned at 
various atomic sites or over clusters of them. The magnetic field is varied 
to control, in a subtle way,  the 
extent of localization and  the location of the 
flat band states in energy space. In addition to this we show that, an 
appropriate tuning of the field can lead to a re-entrant behavior of the 
effective mass of the electron in a band, with a periodic flip in its 
sign.
\end{abstract}
\maketitle
Geometrically frustrated lattices (GFL) supporting flat, dispersionless bands in their 
energy spectrum with macroscopically degenerate eigenstates 
have drawn great interest in recent times~\cite{derzhko}-\cite{flach3}.
Initial interest in antiferromagnetic Heisenberg model on frustrated lattices
~\cite{kikuchi}-\cite{moessner} has evolved into extensive studies of the 
gapped flat band states to gapless chiral modes in graphenes~\cite{yao}, 
in optical lattices of ultracold atoms~\cite{bloch}, waveguide arrays~\cite{christo}, 
or in microcavities having exciton-polaritons~\cite{masumoto}. 
The quenched kinetic energy of an electron in a flat band state (FBS) leads to 
the possibility of achieving strongly correlated electronic states, topologically 
ordered phases, such as the lattice versions of fractional quantum Hall states
~\cite{liu}.
Recently,  
the controlled growth of artificial lattices with complications 
such as in the kagom\'{e} class has added excitement to such studies~\cite{guzman}-
\cite{tamura}.

Spinless fermions are easily trapped in flat bands~\cite{mati}. The non-dispersive 
character of the energy ($E$)- wave vector ($k$) curve implies an infinite 
{\it effective mass} of the electron, leading to practically its {\it immobility}
in the lattice. Such states are therefore strictly localized either on special
 sets of vertices, 
or in a finite cluster of atomic sites spanning finite areas of the underlying lattice. 
Recently it has been shown that, an infinity of such cluster-localized single particle 
states can be exactly constructed even in a class of deterministic 
fractals~\cite{atanu}.
Apart from its interest in direct relation to the study of GFL's, this work 
provides an example where eigenvalues corresponding to localized eigenstates 
in an infinite fractal geometry can be exactly evaluated, a task that is a non-trivial 
one if one remembers that these fractal systems are free from translational invariance 
of any kind.

In this communication we unravel and analyze groups of 
flat, dispersionless energy bands in some tailor made GFL's. The lattices 
display an interesting competition between long range translational 
order along the horizontal ($x-$) axis and an aperiodic growth in the
%%%%%%%%%%%%%%%%%%% FIGURE 1%%%%%%%%%%%%%
\begin{figure}[ht]
{\centering \resizebox*{8.5cm}{5cm}
{\includegraphics{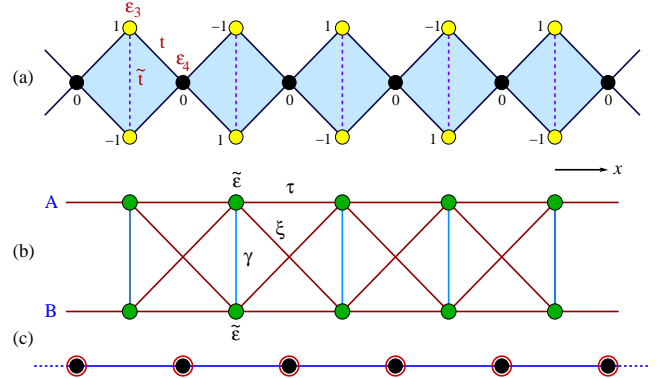}}\par}
\caption{(Color online). (a) A typical realization of an elementary array of 
diamond shaped quadruplets exhibiting a single flat, dispersionless band at $E=0$,   
(b) its renormalization into an effective two-arm ladder network with 
energy dependent potential and nearest and next nearest neighbor interactions 
and, (c) the effective one dimensional chain obtained after decimating out 
the top (yellow colored) vertices.} 
\label{ladder}
\end{figure}
%%%%%%%%%%%%%%%%%%%%%%%%%%%%%%%%%%%%%%%%% 
transverse directions. In each case, the skeleton is an infinite array of 
diamond shaped networks (Fig.~\ref{ladder}(a)) where hierarchical structures,
grown deterministically, are embedded 
in each arm of the diamond. A uniform magnetic flux threads each elementary 
plaquette, as will be illustrated in appropriate cases. The motivation behind the present 
study is two fold.

First, exploiting the self similar growth of the hierarchical structures we 
can implement a real space renormalization group (RSRG) scheme 
to evaluate the {\it pinned}, flat band states (FBS) exactly. In all the cases we discuss 
in this paper the flat band states merge with the localized eigenstates of 
the infinite hierarchical geometry as higher and higher generations
of them are embedded in the arms of the 
underlying diamonds. Thus, once again, the problem of an exact evaluation of 
localized states in a fractal geometry is, at least partially, solved.

Second, and a more important aspect of the problem is 
to look for 
a possible coexistence of a 
multitude of dispersive and non-dispersive energy bands in a periodic array 
of the elementary building blocks. As an arm of the elementary diamond hosts 
a quasi-periodic or a deterministic fractal of sequentially 
increasing generation, the flat, 
dispersive bands are likely get densely packed in an environment of dispersive ones. 
The density of packing may even lead, for a large enough generation of the 
hierarchical structure, to a re-entrant {\it dispersive to 
non-dispersive crossover} in the behavior of the electrons. 
A recent inspiring work by Danieli {\it et al.}~\cite{flach3} has shown that a quasiperiodically
modulated flat band geometry may even allow for a precise engineering of the 
mobility edges. 

In addition to this, we anticipate that, a 
variation in the strength of the magnetic field can lead to a tuning of the 
curvature of the energy dispersion curves. We can thus achieve a 
comprehensive control over the 
group velocity and effective mass of the electron with the help of an external 
field. Knowing that, a deterministic quasiperiodic or fractal geometry 
normally offers a completely fragmented, Cantor set-like energy spectrum, 
this latter study may allow us to control the effective mass of the electron 
using an external agent such as the magnetic field over arbitrarily small scales 
of the energy or, equivalently, the wave vector.

We present here a simple analytical method to detect the 
sharply localized eigenstates that are {\it pinned} on certain atoms or 
atomic clusters in a periodic array of diamonds. The non-dispersive 
character of such states is explicitly worked out. The method is then extended 
to unravel the entire set of such FBS when each arm 
of an elementary diamond  hosts a quasiperiodic lattice  
grown according to a deterministic Fibonacci sequence~\cite{sutherland}, and  
Berker~\cite{griffiths} geometry. Such aperiodic structures with sequentially 
increasing hierarchy are embedded in the diamond's arm. We indeed find that, 
as one gradually increases the degree of aperiodicity in the arms, the 
{\it pinned} FBS in such periodic approximants turn out to be the localized 
eigenstates of the systems in their respective thermodynamic limits. 

In addition, we observe that, with a deterministic aperiodic geometry 
of sufficiently large generation embedded in the arms of a diamond array, the 
interplay of periodicity along the $x$-axis and the aperiodic order in the 
transverse directions, produces a highly complex dispersion pattern. The 
flat, dispersionless bands densely fill the gaps between the dispersive ones, 
giving rise to the possibility of a quasi-continuous crossover between them as the 
aperiodic components grow in hierarchy. 

The magnetic field piercing 
the elementary plaquettes in each case is shown to control the 
group velocity of the electrons, making them more and more immobile as 
the flux $\Phi \rightarrow \Phi_0/2$, where $\Phi_0= hc/e$ is the 
fundamental flux quantum. This is an exemplary case of {\it extreme localization} 
induced by the magnetic flux as discussed by Vidal and co-workers
~\cite{vidal1}-\cite{vidal3}. The magnetic field is shown to flip the 
sign of the effective mass multiple times within a single Brillouin zone -  
a remarkable contrast to the ordinary periodic linear lattice. The 
electron-lattice interaction thus can be fine tuned from outside by 
selective choice of the flux threading a plaquette.

In what follows, we present our results. In section I we work out the 
basic method of analyzing the FBS in an elementary 
diamond array, and compare the result with 
the existing ones. Section II deals with the Fibonacci-diamond chain, and 
in section III we elaborately discuss the fractal-diamond networks. In section IV 
we explicitly discuss the generic diamond loop-array where a magnetic 
field controls the effective mass of the electron making its sign 
periodically flipped. In section V we draw our conclusions.
\section{The Hamiltonian and the pinned eigenstates}
%%%%%%%% THE HAMILTONIAN %%%%%%%%%%%%%%%%%
We refer to Fig.~\ref{ladder}. Spinless, non-interacting electrons are described 
using a tight-binding hamiltonian in the Wannier
basis, viz., 
\begin{equation}
H  = \sum_{i} \epsilon_i c_{i}^{\dagger} c_{i}
+\sum_{\langle ij \rangle} t_{ij} \left[c_{i}^{\dagger} 
c_{j} 
+ h.c. \right] 
\label{hamilton}
\end{equation}
where, $\epsilon_i$ is the on-site potential and can assume 
values equal to $\epsilon_3$ or $\epsilon_4$ depending on the 
site at the vertex having a coordination number $z = 3$ (yellow circles), 
or in the bulk, having 
a coordination number $z = 4$ (black circles). 
Throughout this paper we shall choose $\epsilon_3 = \epsilon_4$ just to 
see the effect of the topology of the lattice alone. However, the 
symbols will be in use to facilitate any discussion.
The nearest 
neighbor hopping integral $t_{ij} = t$ along the arm of the diamond, and 
$t_{ij} = \tilde{t}$ along the diagonal connecting the vertices with 
coordination number two. 
The Schr\"{o}dinger equation, written equivalently in the form of the 
difference equation, 
\begin{equation}
( E - \epsilon_i ) \psi_i = \sum_{j} t_{ij} \psi_{j}
\label{difference}
\end{equation}
allows us to decimate out the ``black" vertices of the diamond network to map the 
original array on to an effective two-arm ladder (Fig.~\ref{ladder}(b)) 
(with arms marked $A$ and $B$) comprising 
identical (green colored) atomic sites with renormalized on-site 
potential $\tilde\epsilon=\epsilon_3 + 2t^2/(E-\epsilon_4)$. The renormalized 
hopping integral along the arm of the ladder now becomes $\tau=t^2/(E-\epsilon_4)$, 
and the inter-arm hopping becomes $\gamma=\tilde{t}+2t^2/(E-\epsilon_4)$. 
The decimation generates a second neighbor hopping (brown line) inside a 
unit plaquette of the ladder and along the diagonal, viz., $\xi=t^2/(E-\epsilon_4)$. 

The difference equation for the ladder network may now be cast 
using $2 \times 2$ matrices, in the form~\cite{sil}:
\begin{widetext}
%%%%%%%%% MATRIX EQUATION %%%%%%%%%%%%%
\begin{eqnarray}
\left [
\left( \begin{array}{cccc}
E & 0 \\
0 & E
\end{array}
\right ) - 
\left( \begin{array}{cccc}
\tilde\epsilon & \gamma \\
\gamma & \tilde\epsilon
\end{array}
\right)
\right ]
\left ( \begin{array}{c}
\psi_{n,A} \\
\psi_{n,B}  
\end{array} \right )
& = & 
\left( \begin{array}{cccc}
\tau & \xi \\ 
\xi & \tau 
\end{array} 
\right)
\left ( \begin{array}{c}
\psi_{n+1,A} \\
\psi_{n+1,B}  
\end{array} \right )
+
\left( \begin{array}{cccc}
\tau & \xi \\                                                  
\xi & \tau
\end{array}
\right)
\left ( \begin{array}{c}
\psi_{n-1,A} \\
\psi_{n-1,B}
\end{array} \right )
\label{eqladder}
\end{eqnarray}
\end{widetext}
%%%%%%%%%%%%%%%%%%%%%%%%%%%%%%%%%%%%%%%%%
It is easy to check that both the `potential matrix' (comprising 
$\tilde\epsilon$ and $\gamma$) and the `hopping matrix' (with $\tau$ and 
$\xi$) commute, and hence can be simultaneously diagonalized by a similarity
 transform. Eq.~\eqref{eqladder} can then be easily decoupled, in a new 
basis defined by $\phi_{n}=\mathbf{M}^{-1}\psi_n$. 
The matrix $\mathbf{M}$  diagonalizes both the `potential' and the 
`hopping' matrices.
The decoupled set of equations are free from any `cross terms' and  
reads, in terms of the original on-site potentials and hopping integrals, as:
\begin{eqnarray}
\left [ E - \left (\epsilon_3 + \tilde{t} + \frac{4t^2}{E-\epsilon_4} \right ) \right ] 
\phi_{n,A} & = & \frac{2t^2}{E-\epsilon_4} ( \phi_{n+1,A} + \phi_{n-1,A} ) \nonumber \\
( E - \epsilon_3 + \tilde{t} ) \phi_{n,B} & = & 0
\label{decouple}
\end{eqnarray}
The first equation represents a periodic array of identical atomic sites 
with renormalized on-site potential $\epsilon_3 + \tilde{t} + 4t^2/(E-\epsilon_4)$, and 
nearest neighbor hopping integral $2t^2/(E-\epsilon_4)$.
The second equation in Eq.~\eqref{decouple} represents (in the new basis) 
an {\it effective} atom, decoupled from its neighbors. The potential of this  
{\it isolated} atomic site is $\epsilon_3 - \tilde{t}$. This leads to an 
eigenfunction with amplitudes {\it pinned} 
at the $z = 3$ vertices as shown in Fig.~\ref{ladder}(a) and the corresponding 
eigenvalue is at $E = \epsilon_3 - \tilde{t}$. 
The amplitude $\psi_i = 0$ at all the $z = 4$ vertices for this special energy. The 
non-zero amplitudes are thus trapped in local clusters ($z = 3$ vertices), 
as discussed in the introduction.
The result obtained by 
Hyrk\"{a}s {\it et al.}~\cite{mati} identifies this state as a flat, non-dispersive 
one~\cite{comment}.

To confirm the non-dispersive character of this {\it pinned} eigenstate, we refer to
Fig.~\ref{ladder}(c), where an effective linear chain of identical atoms is 
obtained by decimating the $\epsilon_3$ vertices. The resulting on-site 
potential and the nearest neighbor hopping integral for this linear lattice is 
given by,
\begin{eqnarray}
\epsilon_0 & = & \epsilon_4 + \frac{4 t^2 (E - \epsilon_3 + \tilde{t})}
{( E - \epsilon_3 )^2 - \tilde{t}^2 } \nonumber \\
t_0 & = & \frac{2 t^2 ( E - \epsilon_3 + \tilde{t})}{( E - \epsilon_3 )^2 - 
\tilde{t}^2 }
\label{linear}
\end{eqnarray}
The linear chain described by Eq.~\eqref{linear} has a dispersion relation, 
\begin{equation}
( E -\epsilon_3 + \tilde{t}) \left [ (E-\epsilon_4) (E-\epsilon_3-\tilde{t}) 
- 8 \tilde{t}^{2} \cos^2 \left(\frac{ka}{2} \right) \right ] = 0
\label{disp}
\end{equation}
In the above equation $a$ is the lattice constant of the effective periodic 
chain.
Eq.~\eqref{disp} clearly indicates a non-dispersive, wave vector-independent,
flat band at $E = \epsilon_3 - \tilde{t}$. This is compliant with the result 
obtained by analyzing the decoupled set of equations Eq.~\eqref{decouple}.
The same set of arguments brings back the FBS at $E=\epsilon_3$ as shown 
in ref~\cite{mati} for $\epsilon_3 = 0$.
%%%%%%%%%%%%%%%%%%%%%%%%%%%%%%%%%%%%%%%%%%%%%%%%%%%%%%%%%%%%%%%%%%%%%%%%%
\section{The Fibonacci-Diamond array}
\begin{figure}[ht]
{\centering \resizebox*{8.5cm}{3cm}
{\includegraphics{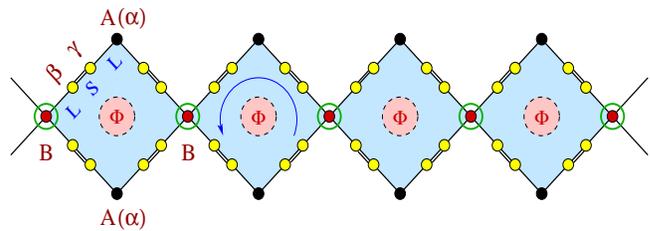}}\par}
\caption{(Color online). (a) A perfectly periodic array of identical
diamond blocks when each arm of a diamond hosts a third generation 
Fibonacci sequence of bonds $L$ and $S$.}
\label{fibonacci}
\end{figure}

We now construct a periodic array of diamond with a Fibonacci segment of two 
different bond lengths $L$ and $S$ embedded in it. 
The linear Fibonacci chain 
grows according to the prescription~\cite{sutherland} $L \rightarrow LS$ and 
$S \rightarrow L$. This generates a quasiperiodic chain with two different 
hopping integrals, viz., $t_L$ and $t_S$ and three kinds of vertices 
$\alpha$ (flanked by two $L$-bonds), $\beta$ ($L$ on the left and $S$ on the 
right), and $\gamma$ (in between an $S-L$ pair). The corresponding on-site
potentials are designated by $\epsilon_\alpha$, $\epsilon_\beta$ and $\epsilon_\gamma$ 
respectively. Fig.~\ref{fibonacci} describes a periodic array of diamonds 
with each arm hosting a third generation Fibonacci segment $LSL$. 
A magnetic flux pierces each diamond. The hopping integrals along the bonds 
$L$ and $S$ pick up Peierls' phases~\cite{vidal1}, and read 
$t_{L}^{f(b)} = t_L \exp(\pm i\theta_L)$ and  
$t_S^{f(b)} = t_S \exp(\pm i\theta_S)$ respectively. Here, 
$\theta_{L(S)}=2 \pi \Phi a_{L(S)}/(F_{n-1}a_L+F_{n-2}a_S)$, $F_n$ being the 
Fibonacci number in the $n$-th generation. The symbols $f$ and $b$ refer to the 
hopping in the so called {\it forward} and {\it backward} directions 
respectively - a consequence 
of the broken time reversal symmetry.
%%%%%%% SECOND LATTICE FIGURE %%%%%%%%%%%%
%%%%%%%%%%%%%%%%%%%%%%%%%%%%%%%%%%%%%%%%% 
In what follows, 
we shall deal with generations ending with the $L$ bond only, that is,
$G_{2n+1}$, $n=1$, $2$, $.........$. This is only for convenience, and does not 
affect the final result as we are interested in the case when $n \rightarrow \infty$. 

As proposed, we embed a $(2n+1)$-th generation Fibonacci segment in each
 arm of the diamond.
 The difference equations Eq.~\eqref{difference} for the three kinds of atomic sites 
are now of the form,
\begin{eqnarray}
(E-\epsilon_\alpha)\psi_j & = & t_{L}e^{i\theta_L}\psi_{j+1} + 
t_{L}e^{-i\theta_L} \psi_{j-1} \nonumber \\
(E-\epsilon_\beta)\psi_j & = & t_{S}e^{i\theta_S}\psi_{j+1} + 
t_{L}e^{-i\theta_L} \psi_{j-1} \nonumber \\
(E-\epsilon_\gamma)\psi_j & = & t_{L}e^{i\theta_L}\psi_{j+1} + 
t_{S}e^{-i\theta_S} \psi_{j-1} 
\label{fibo}
\end{eqnarray}
where, the index $j$ refers to a site of type $\alpha$, $\beta$ or $\gamma$ 
appropriately.

The Fibonacci segment trapped in a diamond arm is now
decimated $n$ times, using the set of Eq.~\eqref{fibo} to reduce 
the geometry into a simple diamond with
%%%%%%%% Dispersion  %%%%%%%%%%
\begin{figure*}[ht]
{\centering \resizebox*{16cm}{14cm}
{\includegraphics{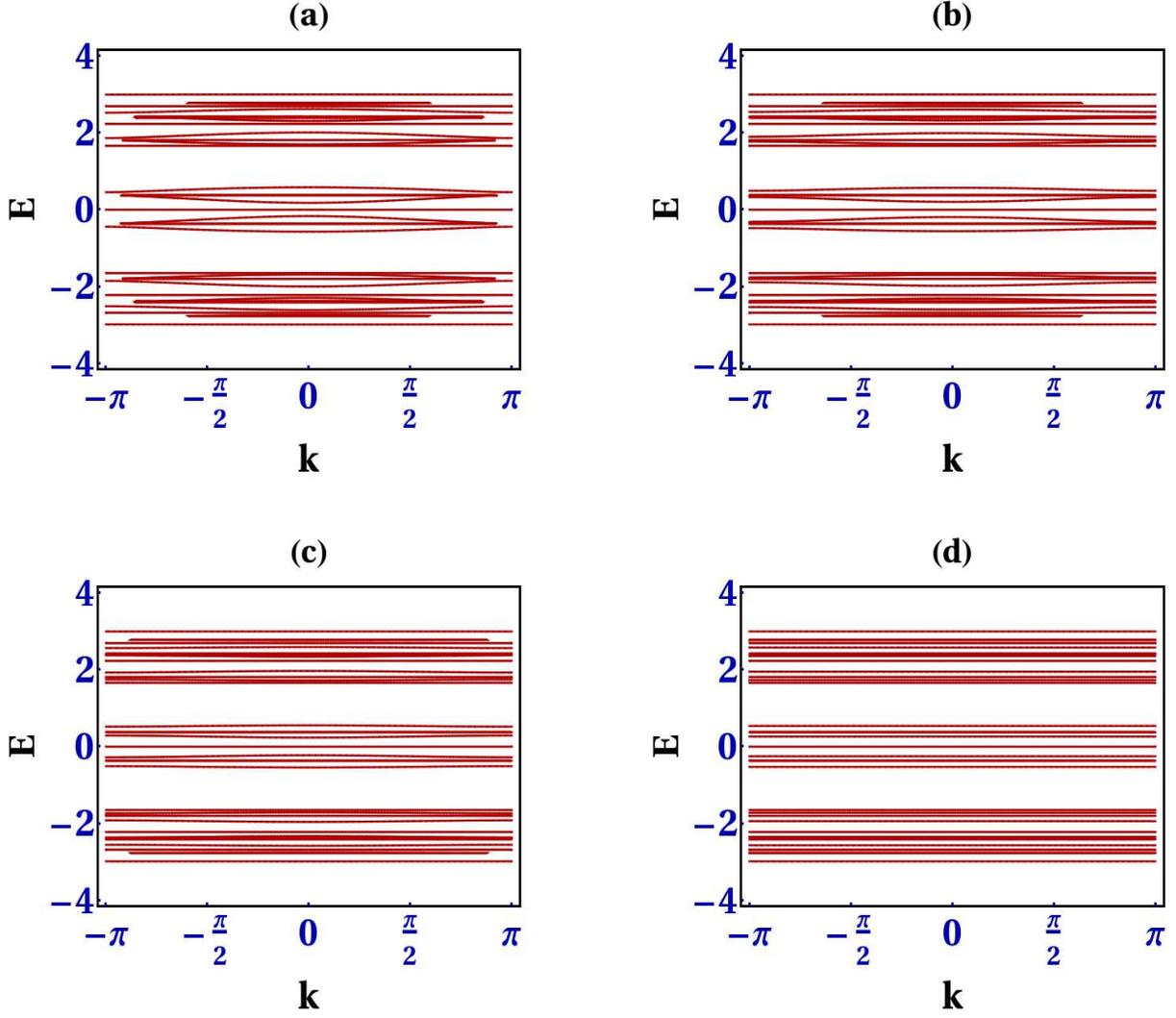}}\par}
\caption{(Color online) Dispersion curves for a 5th generation 
Fibonacci-Diamond array for (a) $\Phi=0$, (b) $\Phi=\Phi_0/4$, 
(c) $\Phi=2 \Phi_0/5$, and (d) $\Phi=\Phi_0/2$. We have chosen 
$\epsilon_\alpha = \epsilon_\beta = \epsilon_\gamma = 0$, 
$t_L=1$ and $t_S=2$. Energy is measured in units of $t_L$.}
\label{dispfibo}
\end{figure*}
%%%%%%%%%%%%%%%%%%%%%%% written upto this %%%%%%%%%%%%%%%%%%%%%%
two kinds of vertices with renormalized on site potentials $\epsilon_{\alpha,n}$, 
(black dots) and $\epsilon_{c,n}$ (red dots encircled with green line). The 
effective hopping integral connecting these two sites is $t_{L,n}^{f(b)}$, depending 
on the sense of traversal.
The recursion relations we exploit during the sequential decimation process 
are given by,
\begin{eqnarray}
\epsilon_{\alpha,n} & = & \epsilon_{\alpha,n-1} + \frac{t_{L,n-1}^f t_{L,n-1}^b
 [2 E - 
(\epsilon_{\beta,n-1} + \epsilon_{\gamma,n-1})]}{\Delta_{n-1}} \nonumber \\ 
\epsilon_{\beta,n} & = & \epsilon_{\alpha,n-1} + \frac{(E-\epsilon_{\beta,n-1}) 
t_{L,n-1}^f t_{L,n-1}^f}{\Delta_{n-1}} +\frac{t_{L,n-1}^f t_{L,n-1}^b}
{E-\epsilon_{\beta,n-1}} \nonumber \\
\epsilon_{\gamma,n} & = & \epsilon_{\gamma,n-1} + \frac{(E-\epsilon_{\gamma,n-1}) 
t_{L,n-1}^f t_{L,n-1}^b}{\Delta_{n-1}} +\frac{t_{S,n-1}^f t_{S,n-1}^b}
{E-\epsilon_{\beta,n-1}} \nonumber \\
\epsilon_{c,n} & = & \epsilon_{\alpha,n-1} + 
2 \left [ \frac{t_{L,n-1}^f t_{L,n-1}^b [2 E -
(\epsilon_{\beta,n-1} + \epsilon_{\gamma,n-1})]}{\Delta_{n-1}} \right ] \nonumber \\ 
t_{L,n}^f & = & \frac{(t_{L,n-1}^f)^2 t_{S,n-1}^f}{\Delta_{n-1}} \nonumber \\
t_{S,n}^f & = & \frac{t_{L,n-1}^f t_{S,n-1}^f}{E-\epsilon_{\beta,n-1}}
\label{recursion1}
\end{eqnarray}
Obviously, $t_{L(S),n}^b = t_{L(S),n}^{f*}$ at any $n$-th stage of renormalization.
The quantity $\Delta_n$ is given by $\Delta_n = 
(E-\epsilon_{\beta,n}) (E-\epsilon_{\gamma,n}) - t_{S,n}^f t_{S,n}^b$.

As the Fibonacci segment is decimated completely following an $n$ step 
execution of the recursion 
relations~\eqref{recursion1}, the Fibonacci-diamond geometry is reduced to a
%%%%%%%%%%%%%%%%%%%%%%%%%%%
\begin{figure*}[ht]
{\centering \resizebox*{16cm}{10cm}{\includegraphics{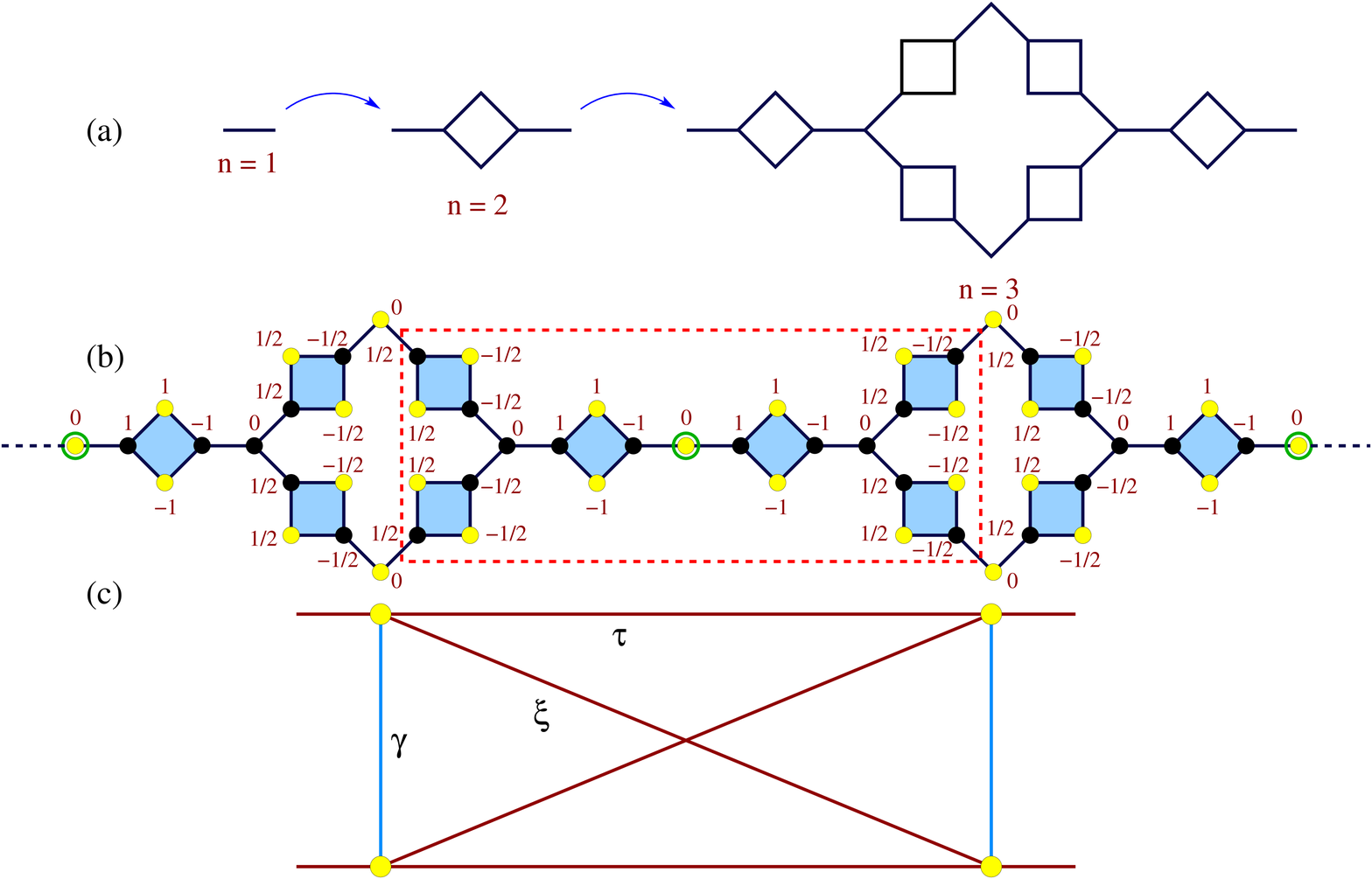}}\par}
\caption{(Color online) (a) Construction of an open Berker-diamond chain.
(b) Distribution of amplitudes $\psi_i$ for a flat band at $E = \epsilon_2 = 0$.
We have set $t = 1$. (c) Conversion of an open Berker-diamond array into a 
two arm ladder network.}
\label{berker}
\end{figure*}
%%%%%%%%%%%%%%%%%%%%%%%%%%%%%%%%%%%%%%%%%%%%%%%%%%%
simple diamond array, as shown in Fig.~\ref{ladder} with $\epsilon_{\alpha,n}$ 
and $\epsilon_{c,n}$ playing the roles of $\epsilon_3$ and $\epsilon_4$
in Eq.~\eqref{linear} respectively. Obviously, $t_{L,n}$ now replaces $t$ and 
$\tilde{t}=0$. The simple analogy reveals that, we have localized eigenstates
pinned at the effective $\alpha$ sites (by virtue of the second equation in 
Eq.~\eqref{decouple}) for all energy eigenvalues obtained by solving the
equation $E - \epsilon_{\alpha,n} = 0$. The corresponding dispersion relation, 
in analogy with Eq.~\eqref{linear} is given by, 
\begin{equation}
(E - \epsilon_{\alpha,n}) \left [ (E - \epsilon_{c,n}) (E - \epsilon_{\alpha,n}) 
- 8 t_{L,n}^2 \cos^2 \left(\frac{ka}{2} \right) \right ] = 0
\label{fibodisp}
\end{equation}
The flat, non-dispersive $k$-independent bands are easily seen to 
originate from the solution of 
the equation $E-\epsilon_{\alpha,n} = 0$.

The solutions of the equation $E=\epsilon_{\alpha,n}$ constitute
the eigenvalue spectrum of a Fibonacci chain~\cite{samar} as 
$n \rightarrow \infty$. The spectrum exhibits a global three subband structure.
Each subband can be scanned over finer scales
of the wave vector to bring out the inherent self similarity and multifractality,
the hallmark of the Fibonacci quasicrystals~\cite{sutherland}. 
In the Fibonacci diamond array, we 
encounter precisely this feature, as
already evident from Fig.~\ref{dispfibo}. 
Each of the three subbands are populated by the dispersive as well as
the non-dispersive FBS. The self similarity of the bands have been checked by 
going over to higher generations, though we refrain from showing these data 
to save space here.

In the limit 
$n \rightarrow \infty$, that is, when a single diamond arm hosts an infinitely 
large Fibonacci segment, the dispersive and non-dispersive bands get 
more and more densely packed, and if one travels along a vertical line at a fixed 
value of the wave vector, a dispersive (non-dispersive) to non-dispersive 
(dispersive) crossover within a single sub-cluster of states is likely to be 
observed. Needless to say, such a crossover can take place over an arbitrarily 
small interval $\Delta k$ of the wave vector. In Fig.~\ref{dispfibo} we have 
shown the bands for a pure {\it transfer model}~\cite{sutherland}, where the 
on-site potentials are set equal to zero, and the quasiperiodic order is 
built in the distribution of the hopping integrals only. The central state at 
$E = 0$ is there for all values of the flux, and as we have checked, for all 
generations. This state belongs to the spectrum of an infinite Fibonacci chain, 
as is well known in the existing literature~\cite{sutherland}.

A magnetic field is found to flatten the dispersive bands, decreasing the
{\it group velocity}, and finally leading to an extreme localization of the 
electronic states. The spectrum then consists entirely of dispersionless, flat 
bands, grouped in triplicate multifractal families. The situation is illustrated 
in Fig.~\ref{dispfibo} where, the panels $(a)$ to $(d)$ sequentially represent 
the grouping of flat and dispersive bands for a diamond array with each arm 
hosting a $5$-th generation Fibonacci chain for $\Phi=0$, $\Phi=\Phi_0/4$, 
$\Phi=2\Phi_0/5$ and $\Phi=\Phi_0/2$ respectively. The gradual 
flattening of the dispersion curves, leading finally to the 
{\it extreme localization} in panel
$(d)$ is self explanatory. 
%%%%%%%%%%%%%%%%%%%%%%%%%%%%%%%%%%%%%%%%%%%%%%%%%%%%%%%
\section{Flat bands in the Berker-diamond arrays}
%%%%%%%%%%%%%%%%%%%%%%%%%%%%%%%%%%%%%%%%%%%%%%%%%%%%%%%
In this section we discuss two cases where each arm of a single
diamond hosts hierarchically grown fractals of the Berker class. The
%%%%%%%%%%%%%%%%%%%%%%%%%%%%%%%%%%%%%%%%%%%%%%%%%%%
essential difference with the earlier case of a Fibonacci diamond is that, 
here, with increasing hierarchy, the lattice grows in an hierarchical
order in the transverse direction relative to the 
arm of a diamond. Periodicity is of course, still maintained 
in the horizontal direction. Each unit is thus an approximant of the 
true, infinite fractal. We begin with an {\it open ended} 
Berker-diamond array whose growth is illustrated in  
Fig.~\ref{berker}(a).

\subsection{The open ended Berker geometry}
The linear periodic array
of approximants of an open ended Berker-diamond array is shown in Fig.~\ref{berker}(b). 
An effective two-arm ladder (Fig.~\ref{berker}(c)) is generated 
by decimating out the vertices trapped inside the 
red dashed boxes. The ladder generated in this way develops, by
construction, a diagonal (second neighbor) hopping (brown line, 
marked $\xi$) which is equal to the effective 
nearest neighbor hopping $\tau$ along an arm. The renormalized on-site potentials and
the hopping integrals of this effective ladder geometry are written, using
the same symbols as in Eq.~\eqref{eqladder} as, 
\begin{eqnarray}
\tilde{\epsilon} & = & \epsilon_{2,n} +\frac{2 \delta_{1,n}t_n^2}{\delta_{2,n}} 
\nonumber \\
\tau & = & \frac{t_n^4}{\delta_{2,n}} \nonumber \\
\gamma & = & \frac{2 \delta_{1,n} t_n^2}{\delta_{2,n}} \nonumber \\
\xi & = & \frac{t_n^4}{\delta_{2,n}}
\label{berklad}
\end{eqnarray}
where, $\delta_{1,n} = (E - \epsilon_{2,n}) (E - \epsilon_{3,n}) - t_{n}^2$, and
$\delta_{2,n} = (E - \epsilon_{3,n}) (\delta_{1,n} - t_{n}^2)$.
In the above set of equations the $z = 2$ and $z = 3$ vertices are 
assigned the on-site potentials $\epsilon_2$ and $\epsilon_3$ respectively, 
though as before, we shall stick to setting $\epsilon_2 = \epsilon_3$.
The difference equation (Eq.~\eqref{eqladder}) for this new ladder
network can again be decoupled, in a new basis, to yield the pair of 
equations, 
\begin{eqnarray}
\left [ E - (\epsilon_{2,n} + \frac{4\delta_{1,n} t_n^2}{\delta_{2,n}}) \right ] \phi_{n,A}
& = & \frac{2t^4}{\delta_2} (\phi_{n+1,A} + \phi_{n-1,A}) \nonumber \\
(E-\epsilon_{2,n})\phi_{n,B} & = & 0
\label{berkdiff}
\end{eqnarray}
Naturally, a group of pinned, FBS are obtained as roots of the equation
$E-\epsilon_{2,n}=0$. 
In Fig.~\ref{berker}(b) we display the distribution of amplitudes 
of the wave function for $E = \epsilon_2 = 0$. The amplitudes are 
non-zero only at the vertices of the smallest squares. One square 
is `separated' from its neighboring ones by a vertex at which the 
amplitude is zero. 
The non-dispersive character of these states can be 
cross-checked, as before, by generating a linear chain connecting 
the green encircled sites in Fig.~\ref{berker}(b). The dispersion relation
becomes, 
%%%%%%%%%%%%%%%%%%%%%%%%%%%%%%%%%%%%%%%%
\begin{equation}
(E - \epsilon_{2,n}) \left [\delta_{1,n}^2 - 4t_n^4 - 2t_n^2 \delta_{1,n} - 
4t_n^4 \cos (ka) \right ] = 0
\label{berkerdisp}
\end{equation}
%%%%%%%%%%%%%%%%%%%%%%%%%%%%%%
\subsection{The closed loop Berker geometry}

\begin{figure}[ht]
{\centering \resizebox*{8.5 cm}{4.4 cm}
{\includegraphics{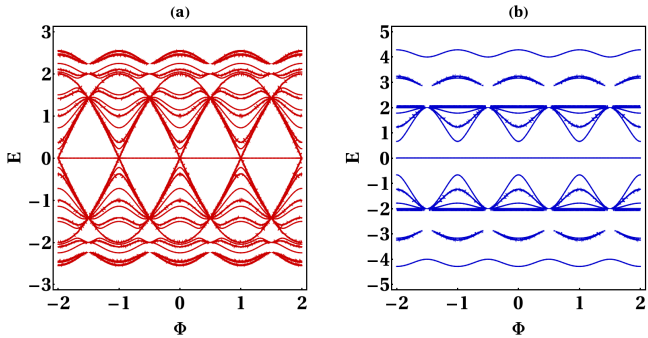}}\par}
\caption{(Color online). Distribution of the flat, non-dispersive 
states only for (a) an open ended Berker-diamond array in its $3$rd 
generation, and (b) a closed loop Berker-diamond array in its $3$rd 
generation. We set $\epsilon_2 = \epsilon_4 = 0$, $t=1$, and energy is measured in units of $t$.} 
\label{flat1}
\end{figure}

In a similar manner, a closed loop Berker-diamond array, shown in
%%%%%%%%%%%%%%%%%%%%%%%%%%%%%%%%%%%%%%%%%%
\begin{figure}[ht]
{\centering \resizebox*{8.5cm}{8 cm}
{\includegraphics{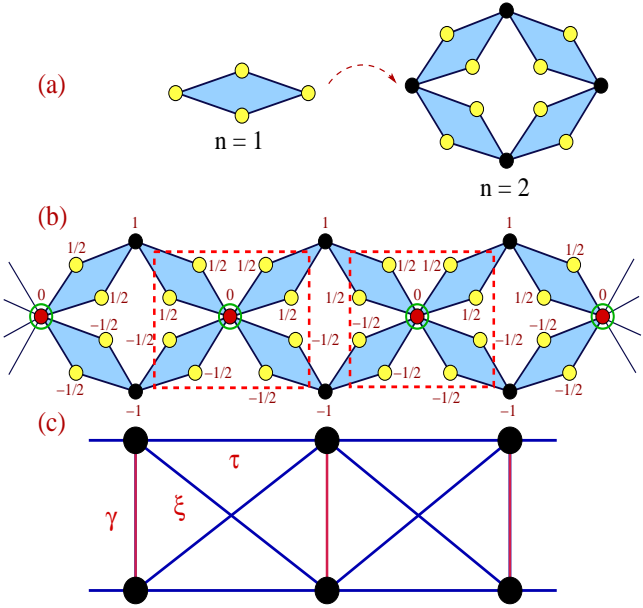}}\par}
\caption{(Color online) (a) Generation and growth of a closed loop
Berker-diamond array. (b) Distribution of amplitudes $\psi_i$ for a flat band at $E = 2$.
 We have set $t = 1$. (c) Conversion of a closed loop Berker-diamond 
array into a 
two arm ladder network.} 
\label{diamond}
\end{figure}
%%%%%%%%%%%%%%%%%%%%%%%%%%%%%%%%%%%%%%%%%% 
Fig.~\ref{diamond}, can be mapped into
an effective ladder (drawn by 
blue lines) by
decimating out the vertices trapped inside the red-dashed boxes. 
There are two kinds of vertices now, viz., with $z = 2$ and $z = 4$.
The on-site potentials bear the symbols $\epsilon_2$ and $\epsilon_4$ 
respectively.
The FBS are extracted by solving the equation $E - \epsilon_{2,n}=0$.
In Fig.~\ref{flat1} we demonstrate the distribution of only the flat band 
states against varying magnetic flux for open-ended (Fig.~\ref{flat1}(a)) 
and closed-loop (Fig.~\ref{flat1}(b)) Berker-diamond arrays. The figure 
brings out the interesting contrast between the two lattices. The FBS for the
%%%%%%%%%%%%%%%%%%%%%%%%%%%%%%%%%%%%%%%%%%%%%%%%
%%%%%%%%%%%%%%%%%%%%%%%%%%%%%%%%%%%%%%%%%%%%%%% 
open ended Berker geometry are grossly divided into two segments above and
below the non-dispersive state at $E=0$. The subgroups of such states touch each other 
at $E = 0$, mimicking a Dirac cone. On the other hand the FBS for the closed 
loop diamond hierarchical array  exhibit global gaps along the energy axis 
as well as for changing values of the magnetic flux. The edge states in both the cases 
are seen to be most affected by the magnetic field, which is likely to lower the
persistent current for such energy regimes.

Before ending this section, it is pertinent to comment that, in all the cases 
discussed so far, the amplitudes of the FBS are confined within a single unit cell 
of the lattice under consideration. Such states thus belong to the $U = 1$ class of 
FBS as per the nomenclature introduced by Danieli {\it et al.}~\cite{flach3}. However, 
the rule of constructing the distribution of amplitudes, as shown as an example in
Fig.~\ref{berker}(b) can be implemented for any arbitrarily large 
generation of the lattice.
This is simple, as any large lattice can be renormalized precisely to the one displayed 
in Fig.~\ref{berker}(b). We need to fix the amplitudes of the flat band wave function to
the values $0$, $\pm 1$ and $\pm 1/2$ on the appropriate vertices of the 
renormalized lattice, exactly in the way 
depicted in Fig.~\ref{flat1}. One can
then easily work in the backward direction to extract the distribution in the bare scale of length
~\cite{atanu}. 
%----------------------------------------------------------

\section{Controlling effective mass with magnetic flux}

The analyses presented in the previous sections lead to an interesting 
prospect. The effective mass of an electron travelling in these kinds of
%%%%%%%%%%%%%%%%%%%%%%%%%%
\begin{figure}[ht]
{\centering \resizebox*{8.5 cm}{8.5 cm}
{\includegraphics{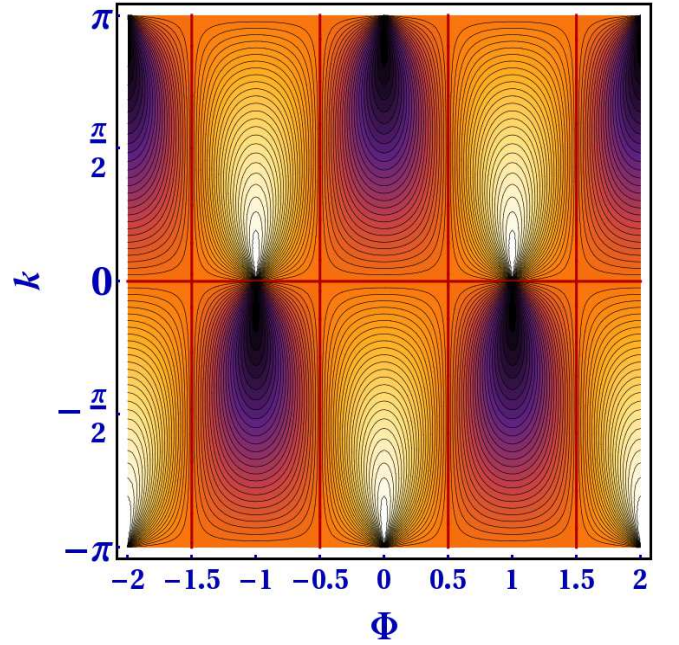}}\par}
\caption{(Color online).  
Fingerprints of the group velocity of the electron in an elementary diamond array. The 
sections between the red lines are filled with a continuous distribution of 
$v_g$. We set $\epsilon_2=\epsilon_4=0$, $t=1$, and display only 
a few contours to help appreciate the discussion in the text.}
\label{group1}
\end{figure}
%%%%%%%%%%%%%%%%%%%%%%%%%% 
arrays can be controlled by tuning the external magnetic field in a non-trivial 
fashion. This result is a generic feature of an array of diamond network, and is thus 
common to the Fibonacci-diamond and Berker-diamond class. 
The microscopic variations are, 
of course sensitive to the specific aperiodic geometry one introduces along the
arm of a diamond. We explain the basics with the help of the simplest 
model of a diamond
array, as depicted in Fig.~\ref{ladder}, but now without the diagonal hopping 
$\tilde{t}$. A uniform magnetic flux $\Phi$ pierces every plaquette. 
Such a geometry can easily be mapped on to an effectively linear chain, 
such as depicted in Fig.~\ref{ladder}(c). The renormalized on-site 
potentials and nearest neighbor hopping integrals on this chain, 
with $\epsilon_2 = \epsilon_4$ are given by, 
\begin{eqnarray}
\epsilon' & = & \epsilon + \frac{4t^2}{E-\epsilon} \nonumber \\
t' & = & \frac{2t^2 \cos(\pi\Phi/\Phi_0)}{E - \epsilon}
\label{chain}
\end{eqnarray}
The dispersive bands are given by, 
\begin{equation}
E = \epsilon \pm 2t \sqrt{1 + \cos \left( \frac{\pi\Phi}{\Phi_0} \right) \cos(ka)}
\label{diamondarray}
\end{equation}
Before presenting the variation of the effective mass with flux, it is essential 
to remind ourselves that, 
as an 
electron travels around a trapped magnetic flux, the wave function picks up a
phase, viz., $\psi \rightarrow \psi_0 \exp\left(i \oint \vec{A}.d\vec{l}\right )$, the 
line integral in the exponent being the flux trapped in the closed loop. The 
magnetic flux here plays a role equivalent to the wave vector~\cite{gefen}. 
One can thus conceive of a $k - \Phi$ diagram which is equivalent to a 
typical $k_x - k_y$ diagram for electrons travelling in a two dimensional 
periodic lattice. The ``Brillouin zone" equivalents are expected to show 
up, across which variations of the group velocity and the effective mass will
take place.

This is precisely the situation as depicted in Fig.~\ref{group1}. 
Every contour displayed corresponds to a fixed value of the group velocity
which exhibits a period equal to $2\Phi_0$. The red lines are the 
equivalents of the Brillouin zone boundaries across which the group 
velocity flips its sign if one moves parallel to the $\Phi$-axis at a fixed 
value of $k$, or vice versa.
This implies that, one can, in principle, 
make an electron {\it retard} without changing its energy by tuning 
the external flux alone. The group velocity is exactly zero along the red lines, 
showing that the eigenfunctions are self-localized around finite clusters 
of sites, making the electronic state a non-dispersive, flat one, as 
we discussed in the beginning.

\begin{figure}[ht]
{\centering \resizebox*{8.5 cm}{8.5 cm}
{\includegraphics{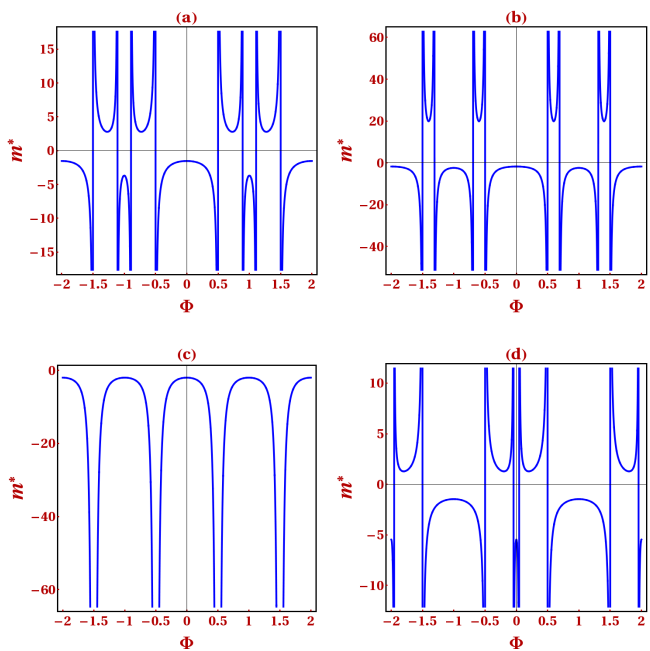}}\par}
\caption{(Color online).  
Effective mass of an electron traversing a simple diamond array plotted 
against the magnetic flux at (a) $ka = \pi/4$, (b) $ka = 2\pi/5$, (c) $ka = \pi/2$, 
and (d) $ka = 5 \pi/6$. We have set $\epsilon = 0$ and $t = 1$ and energy 
is measured in unit of $t$.}
\label{effmass}
\end{figure}

However, a more serious issue crops up as we note that the same 
numerical value of the group velocity may lead to a positive, or a 
negative effective mass of the electron depending on the curvature 
of the dispersive band at those particular values of the group velocity.
This is not apparent from the $k - \Phi$ diagram in Fig.~\ref{group1}, 
but becomes explicit if we look at the expression of the effective mass 
which is given by,   
\begin{equation}
m^{\ast} = \pm \frac{ {\hbar}^2} {a^2 t\left [ \frac{\cos (\frac{\pi\Phi}{\Phi_0}) 
\cos{ka}}
{\sqrt{1 + \cos(\frac{\pi\Phi}{\Phi_0}) \cos{ka}}} + 
\frac{\cos^2 (\frac{\pi\Phi}{\Phi_0})
 \sin^2(ka)}{2[1 + \cos(\frac{\pi\Phi}{\Phi_0})
\cos(ka)]^{3/2}} \right ]}
\label{mass}
\end{equation}
In Fig.~\ref{effmass} we display the variation of the effective mass against
%%%%%%%%%%%%%%%%%%%%%%%%%%%%%%%%%%%%%%%%
%%%%%%%%%%%%%%%%%%%%%%%%%%%%%%%%%%%%%%%%%%%%%%% 
a changing magnetic field (flux) for different values of the wave vector $k$. 
For $0 < ka < \pi/2$ , the 
effective mass displays a {\it re-entrant} behavior, going from a negative value 
below $\Phi < \Phi_0/2$ to a positive value above the half flux quantum through a 
divergence shown precisely at the half flux quantum. The exception to this 
general character takes place at $ka = \pi/2$, which is the first quarter 
of the band. The effective mass remains negative for all values of the 
magnetic flux, diverging at $\Phi = (p+1/2)\Phi_0$, with $p=0$, $\pm 1$, 
$\pm 2$, $.......$
The divergences and crossover in the
sign of the effective mass are repeated in the interval $1/2 \le \Phi/\Phi_0 \le 3/2$.
The variation of $m^{\ast}$ is flux periodic with a period equal to $\Phi/\Phi_0 = 2$.

This discussion points to the obvious implication that the electron-lattice 
interaction can be tuned at will by controlling the magnetic flux from outside.
%%%%%%%%%%%%%%%%%%%%%%%%%%%%%%%%%%%%%%%%%%%%%%%%
The usual idea that the electron is made to behave like a positively charged 
particle in a certain part of the band where its effective mass turns negative 
yields, in this case, a wider variety of phenomenon where 
this can be achieved in a re-entrant 
fashion throughout the band, by tuning an external perturbation, that is, the 
magnetic field. To the best of our knowledge, this simple aspect of such problems 
was not highlighted before.
%%%%%%%%%%%%%%%%%%%%%%%%%%%%%%%%
\section{conclusions}
We have shown how to extract the eigenvalues corresponding to the 
localized, non-dispersive, degenerate flat band eigenstates of an infinite, 
periodic diamond loop array, where aperiodic structures of increasing 
hierarchy grow on each arm of a basic diamond network. A real space 
renormalization group scheme is exploited to unravel a countable infinity 
of such states. The grouping of the dispersive and non-dispersive bands 
in each case resembles the actual band structure of the quasiperiodic or 
fractal lattices considered here, as their generation tends to the respective 
thermodynamic limits. An external magnetic field is shown to be able to control 
the curvature of the dispersive bands, and hence 
the sharpness of localization. The group velocity and the effective 
mass of the electron are thus shown to exhibit a re-entrant behavior inside 
a single Brillouin zone, a fact that turns out to be a consequence of the 
quasi-one dimensionality of the loop structure. 
The scheme is easily extendable to photonic, phononic of magnonic 
excitations, and the flat band eigenstates for an aperiodically grown 
superlattice can thus be worked out. This may throw new challenges 
to experimentalists to engineer the formation and positioning of the 
non-dispersive energy bands in such artificial lattices. A possible 
application in device technology may thus be on the cards. 

\begin{center}
{\bf Acknowledgement}
\end{center}
Atanu Nandy acknowledges financial support from the UGC, India through a 
research fellowship [Award letter no. F.$17$-$81 / 2008$ (SA-I)].

%%%%%%%%%%%%%%%%% Refrerences %%%%%%%%%%%%%%%%%%%%%%%


\begin{thebibliography}{90}
\bibitem{derzhko} O. Derzhko, J. Richter, and A. Honecker, J. Phys.:
Conference Series \textbf{145}, 012059 (2009).
\bibitem{shukla} J. T. Chalker, T. S. Pickles, and P. Shukla, Phys. Rev. B 
\textbf{82}, 104209 (2010).
\bibitem{lopes1} A. A. Lopes and R. G. Dias, Phys. Rev. B \textbf{84}, 085124 (2011).
\bibitem{lopes2} A. A. Lopes. B. A. Z. Ant\'{o}nio, and R. G. Dias, 
Phys. Rev. B \textbf{89}, 235418 (2014).
\bibitem{flach1} D. Leykam, S. Flach, O. Bahat-Treidel, and A. S. Desyatnikov, 
Phys. Rev. B \textbf{88}, 224203 (2013).
\bibitem{mati} M. Hyrk\"{a}s, V. Apaja, and M. Manninen, Phys. Rev. A \textbf{87}, 
023614 (2013).
\bibitem{flach2} S. Flach, D. Leykam, J. D. Bodyfelt, P. Matthies, and 
A. S. Desyatnikov, Europhys. Lett. \textbf{105}, 30001 (2014).
\bibitem{flach3} C. Danieli, J. D. Bodyfelt, and S. Flach, arXiv:1502.06690 (2015).
\bibitem{kikuchi} H. Kikuchi, Y. Fujii, M. Chiba, S. Mitsudo, T. Idehara, 
T. Tonegawa, K. Okamoto, T. Sakai, T. Kuwai, and H. Ohta, Phys. Rev. Lett. 
\textbf{94}, 227201 (2005).
\bibitem{macedo} A. M. S. Mac\^{e}do, M. C. dos Santos, M. D. Coutinho-Filho, 
and C. A. Mac\^{e}do, Phys. Rev. Lett. \textbf{74}, 1851 (1995).
\bibitem{vidal1} J. Vidal, R. Mosseri, and B. Dou\c{c}ot, Phys. Rev. Lett. 
\textbf{81}, 5888 (1998).
\bibitem{vidal2} J. Vidal, B. Dou\c{c}ot, R. Mosseri, and P. Butaud, 
Phys. Rev. Lett. \textbf{85}, 3906 (2000).
\bibitem{vidal3} J. Vidal, P. Butaud, B. Dou\c{c}ot, and R. Mosseri, 
Phys. Rev. B \textbf{64}, 155306 (2001).
\bibitem{moessner} O. Derzhko, J. Richter, A. Honecker, M. Maksymenko, 
and R. Moessner, Phys. Rev. B \textbf{81}, 014421 (2010).
\bibitem{yao} W. Yao, S. A. Yang, and Q. Niu, Phys. Rev. Lett. \textbf{102}, 096801 (2009).
\bibitem{bloch} I. Bloch, J. Dalibard, and W. Zwerger, Rev. Mod. Phys. \textbf{80}, 
885 (2008).
\bibitem{christo} D. N. Christodoulides, F. Lederer, and Y. Silberberg, Nature, 
\textbf{424}, 817 (2003).
\bibitem{masumoto} N. Masumoto, N. Y. Kim, T. Byrnes, K. Kusudo, 
A. L\"{o}ffler, S. H\"{o}fling, A. Forchel, and Y. Yamamoto, 
New. J. Phys. \textbf{14}, 065002 (2012). 
\bibitem{liu} Z. Liu, F. Liu, and Y.-S. Wu, Chin. Phys. B \textbf{23}, 
077308 (2014).
\bibitem{guzman} G.-B. Jo, J. Guzman, C. K. Thomas, P. Hosur, A. Vishwanath, and 
D. M. Stamper-Kurn, Phys. Rev. Lett. \textbf{108}, 045305 (2012).
\bibitem{tamura} K. Shiraishi, H. Tamura, and H. Takayanagi, Appl. Phys. Lett. 
\textbf{78}, 3702 (2001).
\bibitem{atanu} A. Nandy, B. Pal, and A. Chakrabarti, J. Phys.: Condens. Matt. 
\textbf{27}, 125501 (2015).
\bibitem{sutherland} M. Kohmoto, B. Sutherland, and C. Tang, 
Phys. Rev. B \textbf{35}, 1020 (1987).
\bibitem{griffiths} R. Griffiths and M. Kaufman, Phys. Rev. B \textbf{26}, 
5022 (1982).
\bibitem{sil} S. Sil, S. K. Maiti, and A. Chakrabarti, Phys. Rev. Lett. 
\textbf{101}, 076803 (2008).
\bibitem{comment} The eigenstate at $E = \epsilon_3 - \tilde{t}$ is situated 
exactly at the edge of a continuous sub-band of the diamond array, as 
can be easily checked by working out the density of states of this 
periodic system.
\bibitem{samar} S. Chattopadhyay and A. Chakrabarti, Phys. Rev. B 
\textbf{65}, 184204 (2002).
\bibitem{gefen} H.-F. Cheung, Y. Gefen, E. K. Riedel, and 
W.-H. Shih, Phys. Rev. B \textbf{37}, 6050 (1988).
\end{thebibliography}
\end{document}